\documentclass[cernpreprint,texmf,texlive=2011,txfonts,UKenglish,siunitx=false]{atlasdoc}

\pdfoutput=1

\usepackage[biblatex]{atlaspackage}
\usepackage{atlasphysics}
\usepackage{atlasbiblatex}
\AtlasTitle{Search for production of $WW/WZ$ resonances decaying to a lepton, neutrino and jets in $pp$ collisions at $\sqrt{s}=8$~TeV with the ATLAS detector}

\PreprintIdNumber{CERN-PH-EP-2015-044}

\AtlasJournal{Eur. Phys. J. C}

\AtlasAbstract{A search is presented for narrow diboson resonances decaying to $WW$ or $WZ$ in the final state where one $W$ boson decays leptonically (to an electron or a muon plus a neutrino) and the other $W/Z$ boson decays hadronically. The analysis is performed using an integrated luminosity of 20.3~fb$^{-1}$ of $pp$ collisions at $\sqrt{s}=8$~TeV collected by the ATLAS detector at the LHC. No evidence for resonant diboson production is observed, and resonance masses below 760~GeV and 1490~GeV are excluded at 95\% confidence level for the spin-2 Randall--Sundrum bulk graviton $G^*$ with coupling constant of 1.0 and the extended gauge model $W^\prime$ boson respectively.}

\def\antibar#1{\ensuremath{#1\bar{#1}}}

\def\ttbar{\antibar{t}}
\def\pt{\ensuremath{p_{\mathrm{T}}}}
\def\met{\ensuremath{E_{\mathrm{T}}^{\mathrm{miss}}}}

\begin{document}

\maketitle

\section{Introduction}

Several new physics scenarios beyond the Standard Model (SM), such as technicolour~\cite{Eichten2008235,PhysRevD.85.094501,technicolor}, warped extra dimensions~\cite{PhysRevLett.83.3370,PhysRevLett.83.4690,PhysRevD.63.075004}, and grand unified theories~\cite{Altarelli:1989ff}, predict new particles that predominantly  decay to a pair of on-shell gauge bosons. In this paper, a search for such particles in the form of $WW/WZ$ resonances where one $W$ boson decays leptonically ($W\to\ell\nu$ with $\ell=e,\mu$) and the other $W/Z$ boson decays hadronically ($W/Z\to q\bar{q}^\prime/q\bar{q}$, with $q,q^\prime=u,c,d,s\,{\rm or}\,b$) is presented. This search makes use of jet-substructure techniques for highly boosted $W/Z$ bosons decaying hadronically and is optimized to significantly improve the sensitivity to high mass resonances compared to previous searches.

Two benchmark signal models are used to optimize the analysis strategy and interpret the search results. A \mbox{spin-2} Kaluza--Klein (KK) graviton ($G^*$) is used to model a narrow resonance decaying to a $WW$ final state. The KK graviton interpretation is based on an extended Randall--Sundrum model of a warped extra dimension (RS1)~\cite{PhysRevD.76.036006} where the SM fields can propagate into the bulk of the extra dimension. This extended ``bulk'' RS model, referred to as bulk RS hereafter, avoids constraints on the original RS1 model from limits on flavour-changing neutral currents and electroweak precision tests, and has a dimensionless coupling constant $k/\bar{M}_{\rm Pl} \sim 1$, where $k$ is the curvature of the warped extra dimension and $\bar{M}_{\rm Pl}=M_{\rm Pl}/\sqrt{8\pi}$ is the reduced Planck mass. A spin-1 gauge boson ($W^\prime$) of the Sequential Standard Model with modified coupling to $WZ$, also referred to as the extended gauge model (EGM)~\cite{Altarelli:1989ff}, is used to model a narrow resonance that decays to a $WZ$ final state. The EGM introduces $W^\prime$ and $Z^\prime$ bosons with SM couplings to fermions and with the coupling strength of the heavy $W^\prime$ to $WZ$ modified by a mixing factor $\xi = c \times (m_W/m_{W^\prime})^2$ relative to the SM couplings, where $m_W$ and $m_{W^\prime}$ are the pole masses of the $W$ and $W^\prime$ bosons respectively, and $c$ is a coupling scaling factor. In this scenario the partial width of the $W^\prime$ boson scales linearly with $m_{W^\prime}$, leading to a narrow resonance over the accessible mass range. The width of the $W^\prime$ resonance at 1~TeV is approximately 35~GeV.

Searches for these particles in several decay channels have been performed at the Tevatron and the Large Hadron Collider (LHC) and are reported elsewhere~\cite{Khachatryan:2014hpa,Khachatryan:2014gha,PhysRevD.87.112006,tagkey2014223,Aad:2014xka}. Previous results from the ATLAS experiment in the $\ell\ell q\bar{q}$ channel excluded EGM $W^\prime$ bosons with masses up to 1.59~TeV for $WZ$ final states and RS1 gravitons with $k/\bar{M}_{\rm Pl} = 1$ and masses up to 740~GeV for $ZZ$ final states~\cite{Aad:2014xka}. The CMS experiment set limits on the production cross sections of bulk RS gravitons as well as excluded RS1 gravitons with $k/\bar{M}_{\rm Pl}=0.1$ for masses up to 1.2~TeV and $W^\prime$ bosons for masses up to 1.7~TeV~\cite{Khachatryan:2014hpa}.

This analysis is based on $pp$ collision data at a centre-of-mass energy $\sqrt{s}=8$~TeV corresponding to an integrated luminosity of 20.3~fb$^{-1}$ collected by the ATLAS experiment at the LHC.

\section{The ATLAS detector}

The ATLAS detector~\cite{1748-0221-3-08-S08003} is a general-purpose particle detector used to investigate a broad range of physics processes. It includes inner tracking devices surrounded by a superconducting solenoid, electromagnetic and hadronic calorimeters and a muon spectrometer with a toroidal magnetic field. The inner detector (ID) provides precision tracking of charged particles with pseudorapidity $|\eta|<2.5$.\footnote{ATLAS uses a right-handed coordinate system with its origin at the nominal interaction point (IP) in the centre of the detector and the $z$-axis along the beam pipe. The $x$-axis points from the IP to the centre of the LHC ring, and the $y$-axis points upward. Cylindrical coordinates $(r,\phi)$ are used in the transverse plane, $\phi$ being the azimuthal angle around the beam pipe. The pseudorapidity is defined in terms of the polar angle $\theta$ as $\eta=-\ln\tan(\theta/2)$.} The calorimeter system covers the pseudorapidity range $|\eta|<4.9$. It is composed of sampling calorimeters with either liquid argon (LAr) or scintillator tiles as the active media. The muon spectrometer (MS) provides muon identification and measurement for $|\eta|<2.7$. The ATLAS detector has a three-level trigger system to select events for offline analysis.

\section{Monte Carlo samples}

Simulated event samples are used to define the event selection and optimize the analysis. Benchmark signal samples are generated for a range of resonance masses from 300 to 2500~GeV in steps of 100~GeV. The bulk RS $G^*$ signal events are generated with CalcHEP~\cite{Belyaev:2012qa}, using $k/\bar{M}_{\rm Pl}=1.0$, interfaced to {\sc Pythia8}~\cite{PYTHIA8} to model fragmentation and hadronization, and the EGM $W^\prime$ signal is generated using {\sc Pythia8} with $c=1$. The factorization and renormalization scales are set to the generated resonance mass. The CTEQ6L1~\cite{CTEQ6L1} and MSTW2008LO~\cite{MSTW2008LO} parton distribution functions (PDFs) are used for the $G^*$ and $W^\prime$ signal samples respectively. The $W^\prime$ cross section is normalized to a next-to-next-to-leading-order (NNLO) calculation in $\alpha_{\text{s}}$ from {\sc ZWprod}~\cite{Hamberg:1990np}.

Simulated event samples are used to model the shape and normalization of most SM background processes. The main background sources in the analysis arise from $W$ bosons produced in association with jets ($W$+jets), followed by top-quark and multijet production, with smaller contributions from dibosons and $Z$+jets. Production of $W$ and $Z$ bosons in association with up to five jets is simulated using {\sc Sherpa}\,1.4.1~\cite{1126-6708-2009-02-007} with the CT10 PDFs~\cite{PhysRevD.82.074024}, where $b$- and $c$-quarks are treated as massive particles. Samples generated with {\sc MC@NLO}~\cite{1126-6708-2002-06-029} and interfaced to {\sc Herwig}~\cite{HERWIG_1} for hadronization and to {\sc Jimmy}~\cite{Butterworth:1996zw} for the underlying event are used for $\ttbar$ production as well as for single top-quark production in the $s$-channel and the $Wt$ process. The $\ttbar$ cross section is normalized to the calculation at NNLO in QCD including resummation of next-to-next-to-leading logarithmic soft gluon terms with Top++2.0~\cite{Cacciari:2011hy,Beneke2012695,Baernreuther:2012ws,Czakon:2012zr,Czakon:2012pz,Czakon:2013goa,Czakon:2011xx}. Single top-quark production in the $t$-channel is simulated with {\sc AcerMC}~\cite{Kersevan2013919} interfaced to {\sc Pythia6}~\cite{Sjostrand:2006za}. Diboson samples ($WW$, $WZ$ and $ZZ$) are generated with {\sc Herwig} and {\sc Jimmy}.

The effect of multiple $pp$ interactions in the same and neighbouring bunch crossings (pile-up) is included by overlaying minimum-bias events simulated with {\sc Pythia8} on each generated signal and background event. The number of overlaid events is such that the distribution of the average number of interactions per $pp$ bunch crossing in the simulation matches that observed in the data (on average 21 interactions per bunch crossing). The generated samples are processed through the {\sc Geant4}-based detector simulation~\cite{Agostinelli:2002hh,:2010wqa} or a fast simulation using a parameterization of the performance of the calorimeters and {\sc Geant4} for the other parts of the detector~\cite{FastCaloSim}, and the standard ATLAS reconstruction software used for collision data.

\section{Event selection}

Events are required to have a vertex with at least three associated tracks, each with transverse momentum $\pt > 400$~MeV. The primary vertex is chosen to be the reconstructed vertex with the largest track $\sum\pt^2$.

The main physics objects used in this analysis are electrons, muons, jets and missing transverse momentum. Electrons are selected from clusters of energy depositions in the calorimeter that match a track reconstructed in the ID and satisfy ``tight'' identification criteria defined in Ref.~\cite{ElectronID}. The electrons are required to have transverse momentum $\pt>25$~GeV and $|\eta|<2.47$, excluding the transition region between the barrel and endcaps in the LAr calorimeter ($1.37<|\eta|<1.52$). Muons are reconstructed by combining ID and MS tracks that have consistent trajectories and curvatures~\cite{Aad:2014rra}. The muon tracks are required to have $\pt>25$~GeV and $|\eta|<2.5$. In addition, leptons are required to be isolated from other tracks and calorimetric activity. The scalar sum of transverse momenta of tracks with $\pt>1$~GeV within $\Delta R=\sqrt{(\Delta\eta)^2+(\Delta\phi)^2}=0.2$ around the lepton track is required to be less than 15\% of the lepton $\pt$. Similarly, the sum of transverse energy deposits in the calorimeter within a cone of $\Delta R=0.2$, excluding the transverse energy from the lepton and corrected for the expected pile-up contribution, is required to be less than 14\% of the lepton $\pt$. In order to ensure that leptons originate from the interaction point, a requirement of $|d_0|/\sigma_{d_0}<6\,(3.5)$ and $|z_0 \sin\theta|<0.5$~mm is imposed on the electrons (muons), where $d_0$($z_0$) is the transverse (longitudinal) impact parameter of the lepton with respect to the reconstructed primary vertex and $\sigma_{d_0}$ is the uncertainty on the measured $d_0$.

In this analysis, jets are reconstructed from three-dimensional clusters of energy depositions in the calorimeter using two different algorithms. The jet constituents are considered massless. The low-$\pt$ hadronically decaying $W/Z$ candidates are selected by combining the two highest-$\pt$ jets which are constructed by the anti-$k_t$ algorithm~\cite{1126-6708-2008-04-063} with a distance parameter of $R=0.4$. These jets are referred to as small-$R$ jets and denoted by ``$j$'' hereafter. The energy of small-$R$ jets is corrected for losses in passive material, the non-compensating response of the calorimeter, and extra energy due to multiple $pp$ interactions~\cite{JES}. The small-$R$ jets are required to have $\pt>30$~GeV and $|\eta|<2.8$. For jets with $\pt<50$~GeV, the summed scalar $\pt$ of associated tracks from the reconstructed primary vertex is required to be at least 50\% of the summed scalar $\pt$ of all associated tracks. In the pseudorapidity range $|\eta|<2.5$, jets containing hadrons from $b$-quarks are identified using the MV1 $b$-tagging algorithm~\cite{b-tagging} with an efficiency of 70\%, determined from $\ttbar$ simulated events, and with a misidentification rate for selecting light-quark or gluon jets of less than 1\%. 

For high-$\pt$ $W/Z$ bosons, such as the ones from a resonance with mass above 1~TeV, the hadronically decaying $W/Z$ candidates are identified using a single large-$R$ jet, referred to as ``$J$'' hereafter. The Cambridge--Aachen jet clustering algorithm~\cite{Dokshitzer:1997in} with a distance parameter of $R=1.2$ is used. This jet algorithm offers the advantage of allowing the usage of a splitting and filtering algorithm similar to that described in Ref.~\cite{PhysRevLett.100.242001} but optimized for the identification of highly boosted boson decays. To exploit the characteristics of the decay of massive bosons into a light-quark pair, the splitting and filtering algorithm used here does not impose a mass relation between the large-$R$ jet and its subjets~\cite{BosonTaggingNote}. The momentum balance is defined as $\sqrt{y_{\mathrm f}} = \min(\pt^{j1}, \pt^{j2})\Delta R_{12}/m_{12}$, where $\pt^{j1}$ and $\pt^{j2}$ are the transverse momenta of the two leading subjets, $\Delta R_{12}$ is their separation and $m_{12}$ is their invariant mass. To suppress jets from gluon radiation and splitting, $\sqrt{y_{\mathrm f}}$ is required to be greater than 0.45. Furthermore, the large-$R$ jets are required to have $\pt>400$~GeV and $|\eta|<2.0$.

The missing transverse momentum (with magnitude $\met$) is calculated as the negative of the vectorial sum of the transverse momenta of all electrons, muons, and jets, as well as calibrated calorimeter energy clusters within $|\eta|<4.9$ that are not associated with any other objects~\cite{ATLAS_MET}.

The data used were recorded by single-electron and single-muon triggers, which are fully efficient for leptons with $\pt>25$~GeV. The analysis selects events that contain exactly one reconstructed electron or muon matching a lepton trigger candidate, $\met>30$~GeV and no $b$-tagged small-$R$ jets. The transverse momentum of the neutrino from the leptonically decaying $W$ boson is assumed to be equal to the missing transverse momentum. The momentum of the neutrino in the $z$-direction, $p_z$, is obtained by imposing the $W$ boson mass constraint on the lepton and neutrino system, which leads to a quadratic equation. The $p_z$ is defined as either the real component of the complex solution or the smaller in absolute value of the two real solutions.

In order to maximize the sensitivity to resonances with different masses, three different optimized sets of selection criteria are used to classify the events according to the $\pt$ of the leptonically decaying $W$ candidate ($\pt^{\ell\nu}$) and hadronically decaying $W/Z$ candidate ($\pt^{jj}$ or $\pt^{J}$), namely the ``low-$\pt$ resolved region'' (LRR), ``high-$\pt$ resolved region'' (HRR) and ``merged region'' (MR), where the highly boosted $W/Z$ decay products are observed as a single merged jet in the final state. To ensure the orthogonality of the signal regions, events are assigned exclusively to the first region for which the criteria are fulfilled, applying sequentially the MR, HRR, and LRR event selection. The hadronically decaying $W/Z$ candidate is formed by combining the two small-$R$ jets with highest $\pt$ in the resolved regions and its invariant mass $m_{jj}$ is required to be between 65 and 105~GeV. In the LRR (HRR), the event is required to have $\pt^{\ell\nu}>100$~(300)~GeV, $\pt^{jj}>100$~(300)~GeV and $\Delta\phi(j,\met)>1$, where $\Delta\phi(j,\met)$ is the azimuthal angle between the leading jet and the missing transverse momentum. The HRR additionally requires the two leading jets to have $\pt>80$~GeV. In the MR, the large-$R$ jet with the highest $\pt$ is selected as the hadronically decaying $W/Z$ candidate and $\pt^{\ell\nu}>400$~GeV is also imposed. The jet mass of the selected large-$R$ jet ($m_{J}$) is required to be consistent with a $W/Z$ boson mass ($65<m_{J}<105$~GeV) and the azimuthal angle between the jet and the missing transverse momentum, $\Delta\phi(J,\met)$, is required to satisfy $\Delta\phi(J,\met)>1$. The signal acceptance times efficiency after all selection requirements increases from about 5\% at $m_{W^\prime} = 300$~GeV to a plateau of around 25\% for $m_{W^\prime} > 500$~GeV for $W^\prime\to WZ \to \ell\nu q\bar{q}$ with $\ell = e, \mu, \tau$.

\section{Background estimation}

The reconstructed $WW/WZ$ mass, $m_{\ell\nu jj}$ ($m_{\ell\nu J}$), defined as the invariant mass of the $\ell\nu jj$ ($\ell\nu J$) system, is used to distinguish the signal from the background. The background distributions from $W/Z$+jets where $W$ ($Z$) decays leptonically to $\ell\nu$ ($\ell\ell$) considering the three lepton flavors, $\ttbar$, single top-quark and diboson processes are modelled using simulated events. The background shape from multijet production is obtained from an independent data sample that satisfies the signal selection criteria except for the lepton requirement: the electrons are required to satisfy a looser identification criterion (``medium'' in Ref.~\cite{ElectronID}) but not meet the ``tight'' selection criteria; the selected muons are required to satisfy all the selection criteria after inverting the transverse impact parameter significance cut. The contribution from other processes is subtracted from data in the extraction of the multijet background shape.

The background contributions from $\ttbar$, single top-quark and diboson production are normalized to the number of background events predicted by simulation. The $\pt(W)$ distribution in the $W$+jets simulated sample is corrected by comparing it to data in the LRR sidebands defined as $40<m_{jj}<65$~GeV or $105<m_{jj}<200$~GeV. The normalizations of the $W/Z$+jets and multijet background contributions are derived in a control data sample which is obtained by requiring the mass of the hadronic $W/Z$ candidate to be within the $m_J(m_{jj})$ sidebands. They are determined from binned minimum $\chi^2$ fits to the $\met$ distributions in the control data samples corresponding to each signal region and channel separately. The fitted parameters are the normalizations of these two processes. The difference of the $W/Z$+jets normalization from the expected background from simulation ranges between 1\% and 18\%.

The multijet background templates were validated in the electron channel using samples enriched in multijet events, obtained by inverting the $\met$ requirement. The description of the $\ttbar$ background in simulation was validated in a sample dominated by top-pair events by requiring at least one $b$-tagged small-$R$ jet. Good agreement within uncertainties is observed between data and expectation in these validation regions.

\section{Systematic uncertainties}

The main systematic uncertainty on the background estimation is the uncertainty on the normalization of $W/Z$+jets background obtained from the fit described above. This uncertainty is 3--4\% in the LRR and HRR, and 13--19\% in the MR. An uncertainty on the shape of the $W/Z$+jets background is obtained in the LRR by comparing data and simulation in the $m_{jj}$ sidebands, leading to an approximately 5\% uncertainty for $m_{\ell\nu jj}<600$~GeV. Due to the low numbers of data events in the sidebands for the HRR and MR, the $W$+jets shape uncertainty in these regions is evaluated by comparing a sample of simulated events from {\sc Sherpa} with a sample of simulated events from {\sc Alpgen}~\cite{ALPGEN} interfaced to {\sc Pythia6}. The uncertainty in the shape of the $\ttbar$ mass distribution is estimated by comparing a sample from {\sc MC@NLO} interfaced to {\sc Herwig} with a sample from {\sc Powheg}~\cite{Nason:2004rx,Frixione:2007vw,Alioli:2010xd} interfaced to {\sc Pythia6}. The uncertainty on the shape of the multijet background is evaluated by using alternative templates obtained by removing the calorimeter-based lepton isolation cuts. For the remaining background processes, detector-related uncertainties from the small-$R$ jet energy scale and resolution, large-$R$ jet energy and mass scale, lepton reconstruction and identification efficiencies, lepton momentum scales and resolutions, and missing transverse momentum were considered when evaluating possible systematic effects on the shape or normalization of the background estimation and are found to have a minor impact. The large-$R$ jet energy and mass scale uncertainties are evaluated by comparing the ratio of calorimeter-based to track-based measurements in dijet data and simulation, and are validated by in-situ data of high-$\pt$ $W$ production in association with jets.

The dominant uncertainty on the signal arises from initial- and final-state radiation modelling in {\sc Pythia} and is less than 12\% (6\%) for $G^*$ ($W^\prime$). Uncertainties due to the choice of PDFs are below 1\%.

The uncertainty on the integrated luminosity is $\pm 2.8$\%. It is determined, following the same methodology as that detailed in Ref.~\cite{ATLAS_Luminosity2012}, from a calibration of the luminosity scale derived from beam-separation scans performed in November 2012.

\section{Results and interpretation}

\begin{figure}
\centering
\includegraphics[width=0.47\textwidth]{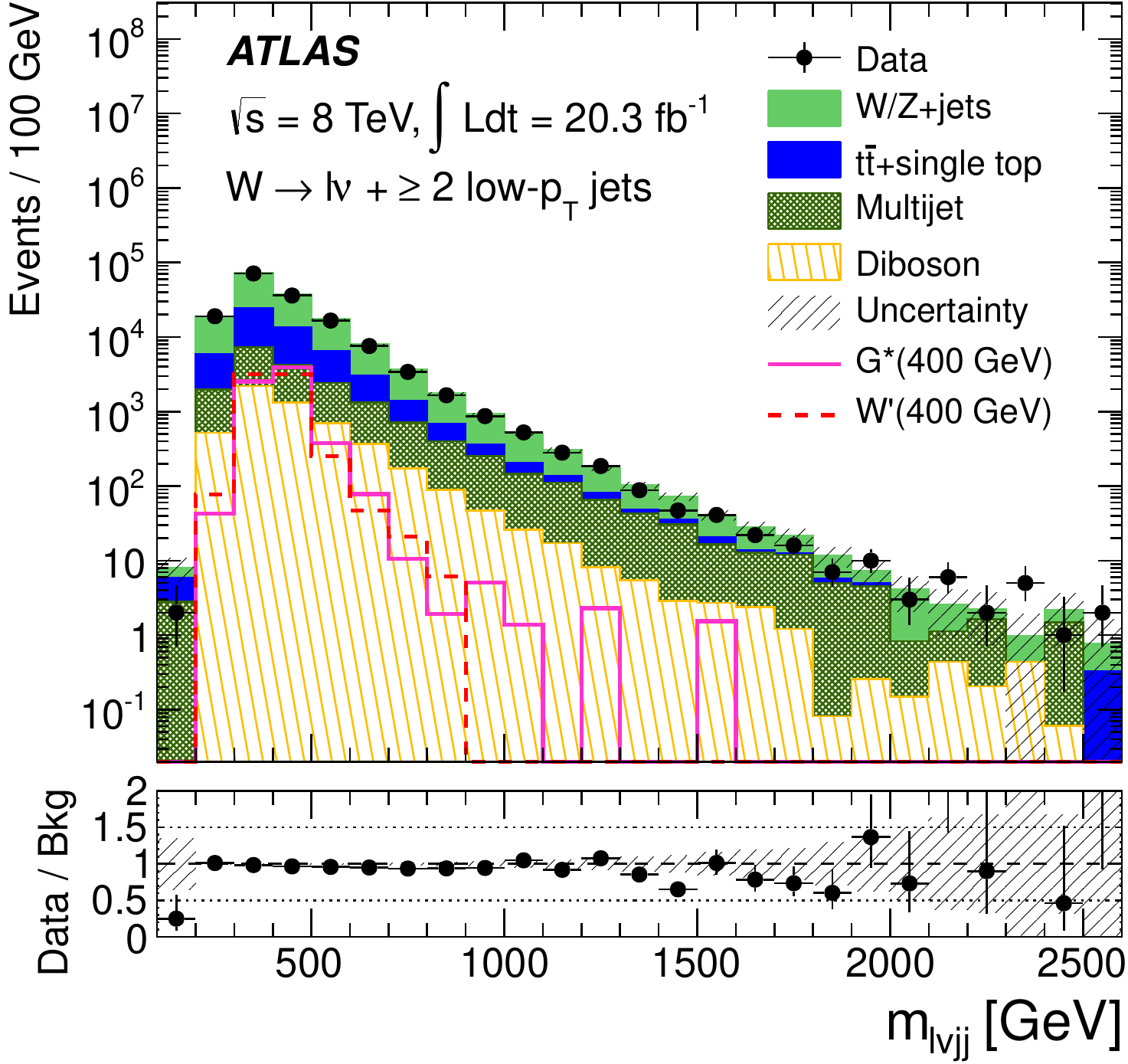}\qquad
\includegraphics[width=0.47\textwidth]{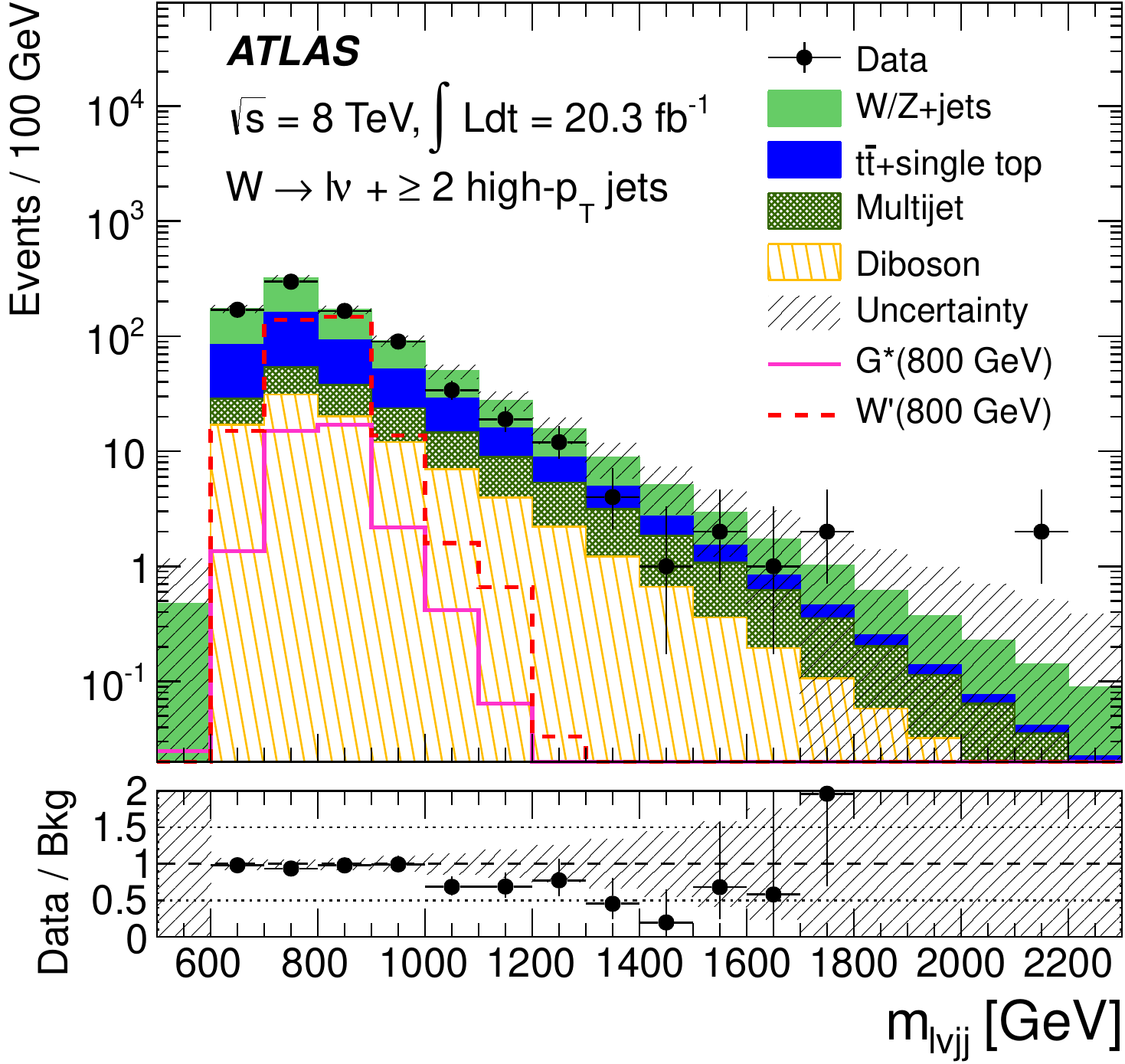}\\\vspace{5mm}
\includegraphics[width=0.47\textwidth]{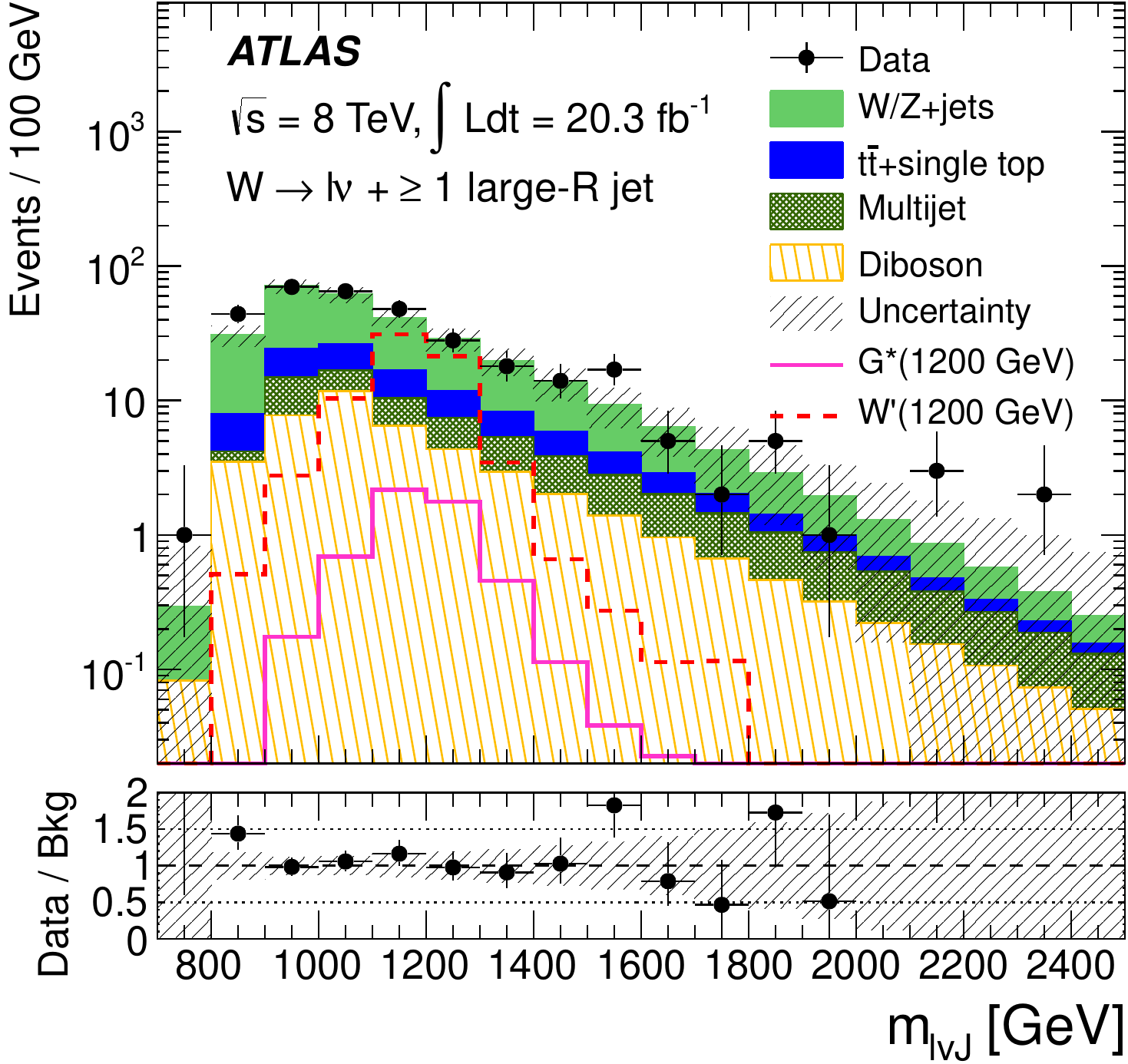}
\caption{Reconstructed mass distributions in data and the predicted backgrounds in the three kinematic regions referred to in the text as the low-$\pt$ resolved region (top left), high-$\pt$ resolved region (top right) and merged region (bottom). $G^*$ and $W^\prime$ signal hypotheses of masses 400, 800 and 1200~GeV are also shown. The band denotes the statistical and systematic uncertainty on the background before the fit to the data. The lower panels show the ratio of data to the SM background estimate.}
\label{fig:mass}
\end{figure}

\begin{figure}
\centering
\includegraphics[width=0.7\textwidth]{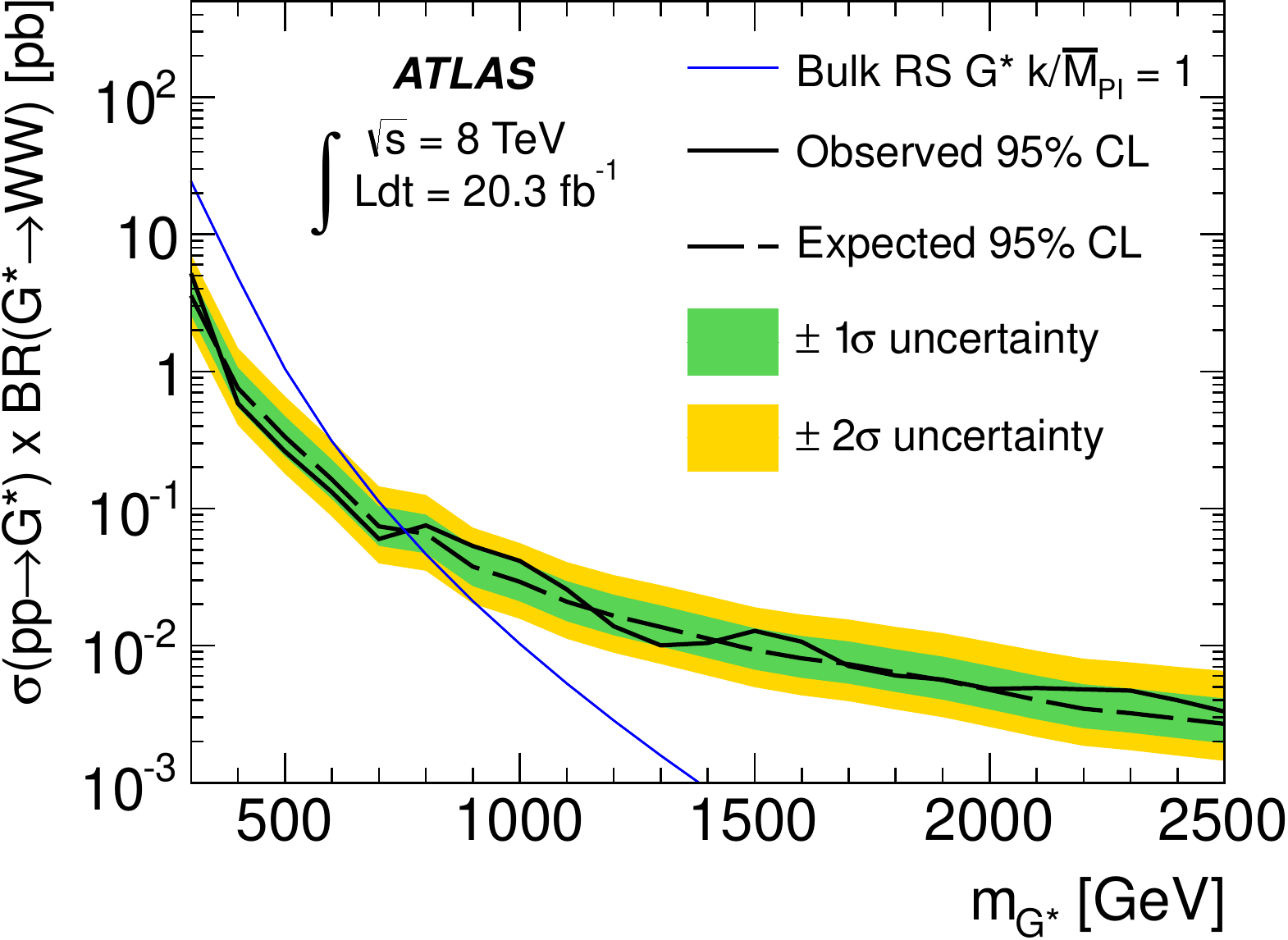}\\\vspace{5mm}
\includegraphics[width=0.7\textwidth]{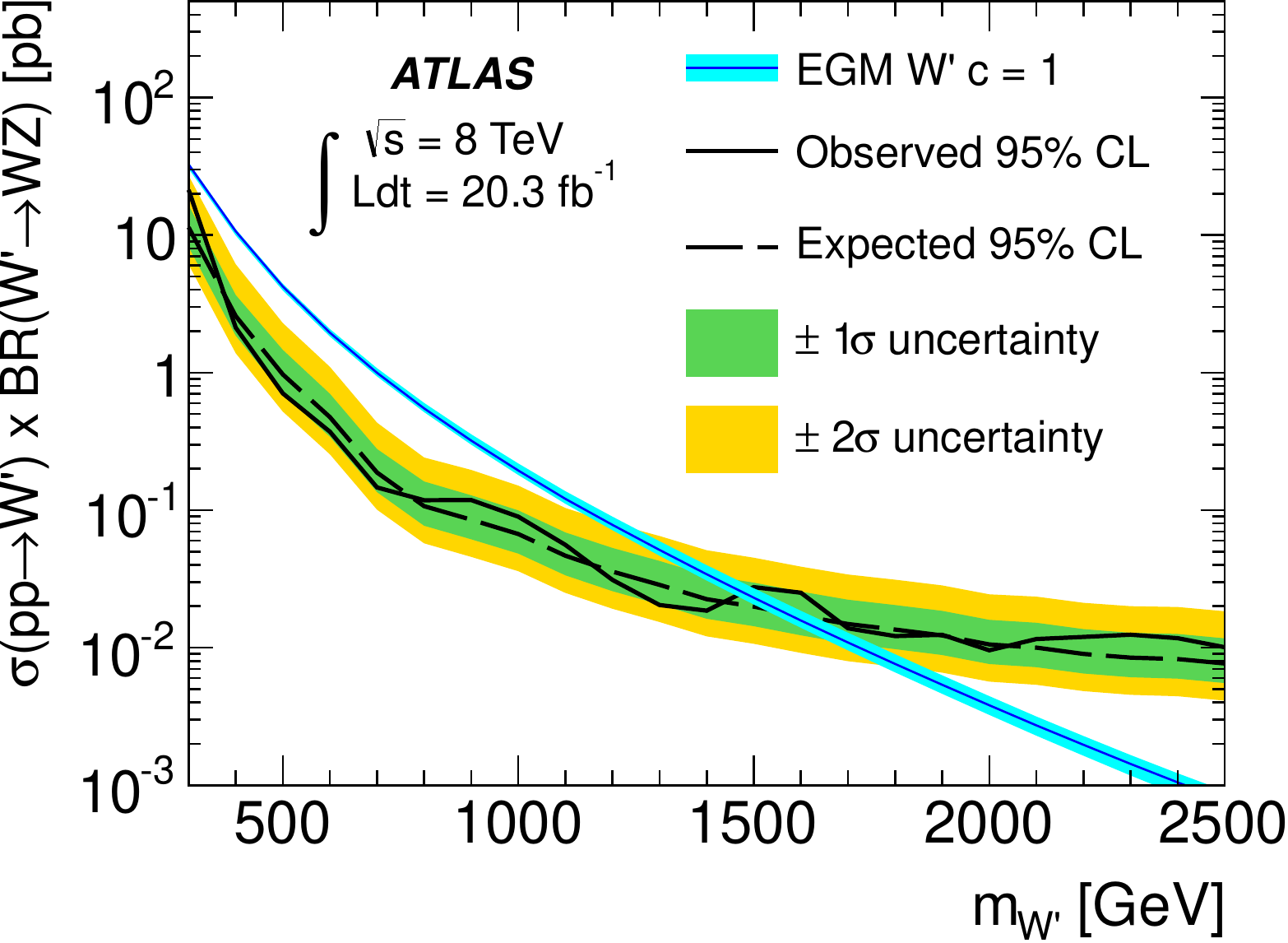}
\caption{Observed and expected 95\% CL upper limits on the cross section times branching fraction as a function of the resonance pole mass for the $G^*$ (top) and EGM $W^\prime$ (bottom). The LO (NNLO) theoretical cross section for the $G^*$ (EGM $W^\prime$) production is also shown. The inner and outer bands around the expected limits represent $\pm 1\sigma$ and $\pm 2\sigma$ variations respectively. The band around the $W^\prime$ cross section corresponds to the NNLO theory uncertainty.}
\label{fig:limit}
\end{figure}

Table~\ref{tab:yield} shows the number of events predicted and observed in each signal region. The reconstructed $m_{\ell\nu jj}$ $(m_{\ell\nu J})$ distributions for data and predicted background events as well as selected benchmark signal models in the three signal regions are shown in Fig.~\ref{fig:mass} for the combined electron and muon channels. Good agreement is observed between the data and the background prediction. In the absence of a significant excess, the result is interpreted as 95\% confidence level (CL) upper limits on the production cross section times branching ratio for the $G^*$ and $W^\prime$ models. These upper limits are determined with the CL$_{\text{s}}$ modified frequentist formalism~\cite{CLs_2} with a profile-likelihood test statistic~\cite{cowan2011}. The test statistic is evaluated with a maximum-likelihood fit of signal models and background predictions to the reconstructed $m_{\ell\nu jj} (m_{\ell\nu J})$ spectra. Systematic uncertainties are taken into account as nuisance parameters with Gaussian sampling distributions. For each source of systematic uncertainty, the correlations across bins and between different kinematic regions, as well as those between signal and background, are taken into account. The likelihood fit is performed for signal pole masses between 300 and 800~GeV for the LRR, 600--1000~GeV for the HRR and 800--2000~GeV for the MR. Overlapping regions are fit simultaneously. Figure~\ref{fig:limit} shows 95\% CL upper limits on the production cross section multiplied by the branching fraction into $WW$ ($WZ$) for the bulk RS $G^*$ (EGM $W^\prime$) as a function of the resonance pole mass. The theoretical predictions for the EGM $W^\prime$ with a scale factor $c=1$ and the bulk RS $G^*$ with coupling constant $k/\bar{M}_{\rm Pl}=1$, shown in the figure, allow observed lower mass limits of 1490~GeV for the $W^\prime$ and 760~GeV for the $G^*$ to be extracted.

\begin{table}
\hfill{}
\caption{Event yields in signal regions for data, predicted background contributions, and $G^*$ and $W^\prime$ signals. Errors are shown before the fit to the data. The errors on the total background and total signal correspond to the full statistical and systematic uncertainty, while the errors on each background component include the full systematic uncertainty only. The $G^*$ and $W^\prime$ signal hypotheses correspond to resonance masses of 400, 800 and 1200~GeV for the LRR, HRR, and MR selections, respectively.}
\vspace{5mm}
\label{tab:yield}
\begin{center}
\begin{tabular}{lr@{$\, \pm \,$}lr@{$\, \pm \,$}lr@{$\, \pm \,$}l}
\hline\hline
Sample                  & \multicolumn{2}{c}{LRR} & \multicolumn{2}{c}{HRR} & \multicolumn{2}{c}{MR} \\
\hline
$W/Z$\,+\,jets          & 104800 & 1600 & 415 & 10 & 180   & 20   \\
$\ttbar$\,+\,single top &  37700 & 1600 & 271 & 13 &  42   &  7   \\
Multijet                &  13500 &  500 &  84 &  9 &  29.3 &  2.9 \\
Diboson                 &   5500 &  270 &  96 &  6 &  43   &  7   \\
\hline
Total                   & 161500 & 2300 & 870 & 40 & 295   & 22   \\
\hline
Data                    & \multicolumn{2}{c}{157837} & \multicolumn{2}{c}{801} & \multicolumn{2}{c}{323} \\
\hline
$G^*$ signal            &   7000 &  500 &  36 &  6 &   5.5 &  2.3 \\
$W^\prime$ signal       &   6800 &  600 & 318 & 21 &  70   &  4   \\
\hline\hline
\end{tabular}
\end{center}
\end{table}

\section{Summary}

A search for $WW$ and $WZ$ resonances decaying to a lepton, neutrino and jets is presented in this paper. The search is performed using an integrated luminosity of 20.3~fb$^{-1}$ of $pp$ collisions at $\sqrt{s}=8$~TeV collected by the ATLAS detector at the LHC. A set of event selections for bulk RS $G^*$ and EGM $W^\prime$ boson signal is derived using simulated events and applied to the data. No evidence for resonant diboson production is observed and 95\% CL upper limits on the production cross section times branching fraction of $G^*$ and $W^\prime$ are determined. Resonance masses below 760~GeV are excluded for the spin-2 RS graviton $G^*$ and masses below 1490~GeV are excluded for the spin-1 EGM $W^\prime$ boson at 95\% CL. The analysis also sets the most stringent limits to date on the production cross section for $W^\prime$-like resonances decaying to $WZ$ with masses around 2~TeV, where $\sigma(pp\to W^\prime)\times\text{BR}(W^\prime\to WZ)$ values of 9.6 fb are excluded. The results represent a significant improvement over previously reported limits~\cite{PhysRevD.87.112006} in the same final state due to an increased data set size and the development of new techniques to analyse highly boosted bosons that decay hadronically. 

\section*{Acknowledgements}

We thank CERN for the very successful operation of the LHC, as well as the
support staff from our institutions without whom ATLAS could not be
operated efficiently.

We acknowledge the support of ANPCyT, Argentina; YerPhI, Armenia; ARC,
Australia; BMWFW and FWF, Austria; ANAS, Azerbaijan; SSTC, Belarus; CNPq and FAPESP,
Brazil; NSERC, NRC and CFI, Canada; CERN; CONICYT, Chile; CAS, MOST and NSFC,
China; COLCIENCIAS, Colombia; MSMT CR, MPO CR and VSC CR, Czech Republic;
DNRF, DNSRC and Lundbeck Foundation, Denmark; EPLANET, ERC and NSRF, European Union;
IN2P3-CNRS, CEA-DSM/IRFU, France; GNSF, Georgia; BMBF, DFG, HGF, MPG and AvH
Foundation, Germany; GSRT and NSRF, Greece; RGC, Hong Kong SAR, China; ISF, MINERVA, GIF, I-CORE and Benoziyo Center, Israel; INFN, Italy; MEXT and JSPS, Japan; CNRST, Morocco; FOM and NWO, Netherlands; BRF and RCN, Norway; MNiSW and NCN, Poland; GRICES and FCT, Portugal; MNE/IFA, Romania; MES of Russia and NRC KI, Russian Federation; JINR; MSTD,
Serbia; MSSR, Slovakia; ARRS and MIZ\v{S}, Slovenia; DST/NRF, South Africa;
MINECO, Spain; SRC and Wallenberg Foundation, Sweden; SER, SNSF and Cantons of
Bern and Geneva, Switzerland; NSC, Taiwan; TAEK, Turkey; STFC, the Royal
Society and Leverhulme Trust, United Kingdom; DOE and NSF, United States of
America.

The crucial computing support from all WLCG partners is acknowledged
gratefully, in particular from CERN and the ATLAS Tier-1 facilities at
TRIUMF (Canada), NDGF (Denmark, Norway, Sweden), CC-IN2P3 (France),
KIT/GridKA (Germany), INFN-CNAF (Italy), NL-T1 (Netherlands), PIC (Spain),
ASGC (Taiwan), RAL (UK) and BNL (USA) and in the Tier-2 facilities
worldwide.

\printbibliography

\clearpage
\begin{flushleft}
{\Large The ATLAS Collaboration}

\bigskip

G.~Aad$^{\rm 85}$,
B.~Abbott$^{\rm 113}$,
J.~Abdallah$^{\rm 152}$,
S.~Abdel~Khalek$^{\rm 117}$,
O.~Abdinov$^{\rm 11}$,
R.~Aben$^{\rm 107}$,
B.~Abi$^{\rm 114}$,
M.~Abolins$^{\rm 90}$,
O.S.~AbouZeid$^{\rm 159}$,
H.~Abramowicz$^{\rm 154}$
% ...
\bigskip
\\
% ...
$^{11}$ Institute of Physics, Azerbaijan Academy of Sciences, Baku, Azerbaijan\\
% ...
$^{85}$ CPPM, Aix-Marseille Universit{\'e} and CNRS/IN2P3, Marseille, France\\
% ...
$^{90}$ Department of Physics and Astronomy, Michigan State University, East Lansing MI, United States of America\\
% ...
$^{107}$ Nikhef National Institute for Subatomic Physics and University of Amsterdam, Amsterdam, Netherlands\\
% ...
$^{113}$ Homer L. Dodge Department of Physics and Astronomy, University of Oklahoma, Norman OK, United States of America\\
$^{114}$ Department of Physics, Oklahoma State University, Stillwater OK, United States of America\\
% ...
$^{117}$ LAL, Universit{\'e} Paris-Sud and CNRS/IN2P3, Orsay, France\\
% ...
$^{152}$ Institute of Physics, Academia Sinica, Taipei, Taiwan\\
% ...
$^{154}$ Raymond and Beverly Sackler School of Physics and Astronomy, Tel Aviv University, Tel Aviv, Israel\\
% ...
$^{159}$ Department of Physics, University of Toronto, Toronto ON, Canada\\
% ...
\end{flushleft}


\end{document}